\documentclass[aps,pra,twocolumn,showpacs,groupedaddress]{revtex4-2}  
\usepackage{amsmath}
\usepackage{tipa}
\usepackage{bbm}
\usepackage{txfonts}
\usepackage{graphicx}
\usepackage{dcolumn}
\usepackage{bm}
\usepackage{amssymb}
\usepackage{latexsym}
\usepackage{color}
\usepackage{ulem}
\begin{document}


\title{Interaction effects of pseudospin based magnetic monopoles and kinks in a doped dipolar superlattice gas}
\author{Xiang Gao$^1$} 
\author{Shao-Jun Li$^1$}
\author{Shou-Long Chen$^1$}
\author{Xue-Ting Fang$^1$}
\author{Qian-Ru Zhu$^1$}
\author{Xing Deng$^1$}

\author{Lushuai Cao$^1$}\email[E-mail: ]{lushuai_cao@hust.edu.cn}
\author{Peter Schmelcher$^{2,3}$}
\author{Zhong-Kun Hu$^1$}\email[E-mail: ]{zkhu@hust.edu.cn}
\affiliation{$^1$MOE Key Laboratory of Fundamental Physical Quantities Measurement $\& $ Hubei Key Laboratory of Gravitation and Quantum Physics, 
PGMF and School of Physics, Huazhong University of Science and Technology, Wuhan 430074, P. R. China \\
$^2$Zentrum für optische Quantentechnologien, Universität Hamburg, Luruper Chaussee 149, 22761 Hamburg, Germany \\
$^3$The Hamburg Centre for Ultrafast Imaging, Universität Hamburg, Luruper Chaussee 149, 22761 Hamburg, Germany
}
\date{\today}

\begin{abstract}
Magnetic monopoles and kinks are topological excitations extensively investigated in quantum spin systems, 
but usually they are studied in different setups. We explore the conditions for the coexistence and the interaction
effects of these quasiparticles in the pseudospin chain of the atomic dipolar superlattice gas. In this chain, the
magnetic kink is the intrinsic quasiparticle, and the particle/hole defect takes over the role of the north/south
magnetic monopole, exerting monopolar magnetic fields to neighboring spins. A binding effect between the monopole 
and kink is revealed, which renormalizes the dispersion of the kink. The corresponding dynamical antibinding process
is observed and arises due to the kink-antikink annihilation. The rich interaction effects of the two quasiparticles
could stimulate corresponding investigations in bulk spin systems.

\end{abstract}

\pacs{}

\maketitle

\setlength{\arraycolsep}{0.8pt}

\section{INTRODUCTION} \label{section:I}

Quantum spin systems possess various topological excitations, such as the magnetic kink
\cite{Wres95,Rutk10}, the spinon \cite{Caux13,Becca19}, the skyrmion \cite{Zaharko20}, the Majorana mode
\cite{Surendran19}, as well as the magnetic monopole \cite{Sondhi08,Hold09,Bram09,Perry09}. These quasiparticles
possess rich magnetic properties and endow the spin systems with potential applications, such as functional spintronic
devices \cite{Allwood05,Stuart08,Yan12}. The interaction between different quasiparticles is of particular importance,
since it not only enriches the dynamical properties of the spin systems, but also provides efficient manipulation tools
for corresponding applications. The coexistence and interaction effects between different quasiparticles, such as
magnon and spinon \cite{Ke20,Starykh20}, kink and magnon \cite{Kirschner04}, as well as magnon and skyrmion
\cite{Nagaosa14,Garst14,Yanpeng21}, have been extensively investigated.
Appealing coupling effects have been revealed between these quasiparticles, which manifests the demand
for the investigation of so far unknown interaction effects of magnetic quasiparticles, specifically 
the monopole and the kink.  This could be explored both
in condensed matter spin systems and/or in pseudospin systems emulated with e.g. ultracold atomic gases.

Ultracold atomic gases have become one of the major platforms for quantum simulation \cite{Piet0902,Piet0906,Greiner11,Gross13,Ha14,Ha15,Cao15,Ketterle16,Zinner17,Santos17,Cao17,Ketterle17,Gross17,Greiner17,Spielman18,Sadeghpour18,Koep19,Greiner19,Jayadev20,Gorshkov19,Monroe21}, 
thanks to the rich degrees of freedom of the atomic gas to construct the target Hilbert space and the tunability to engineer the demanded Hamiltonians. 
Concerning quantum simulation of spin systems, the target spin degree of freedom can be modeled by the atomic species \cite{Zinner17,Santos17}, 
the atomic internal states \cite{Gross13,Greiner17,Sadeghpour18,Koep19,Greiner19,Jayadev20,Gorshkov19,Monroe21}, as well as the spatial modes of lattice atoms, such as the occupation states in tilted lattices \cite{Greiner11} or superlattices \cite{Cao15,Ketterle16,Cao17,Ketterle17}. 
The effective interaction such as spin-spin interactions \cite{Sadeghpour18} and the spin-orbital couplings \cite{Ketterle16,Ketterle17} have also been engineered. Various magnetic quasiparticles have been simulated, such as the magnons \cite{Gross13}, spinons \cite{Jayadev20}, spin/magnetic polarons \cite{Greiner19,Koep19} and the magnetic kinks \cite{Cao15,Gorshkov19,Monroe21}. Particularly, magnetic monopoles of different types have been both theoretically and experimentally implemented for atomic Bose-Einstein condensates \cite{Piet0902,Piet0906,Ha14,Ha15,Spielman18}, with the generation, the dynamical properties and interaction effects investigated.

The magnetic monopole generated in ultracold atomic \cite{Piet0902,Piet0906,Ha14,Ha15,Spielman18}
and condensed matter systems \cite{Sondhi08,Hold09,Bram09,Perry09} are mainly embedded in the superfluid and the spin
ice phases, respectively, in which the excitation condition and dynamical properties
of the monopoles have been extensively investigated.
These magnetic phases, however, can hardly sustain the coexistence of the monopole with other magnetic quasiparticles,
and hinder the investigation of their coupling effects.
In this letter we propose a new quantum simulation scheme, which generates the monopole on the ferromagnetic
host background and enables the coexistence and interaction between the monopole and the intrinsic ferromagnetic
quasiparticle, i.e. the kink. Our simulation scheme adapts and generalizes 
the pseudospin mapping of ultracold atoms in a double-well superlattice, which has been exploited to simulate the 
spin-orbit coupling \cite{Ketterle16} and the corresponding supersolid-like phase \cite{Ketterle17}, 
as well as magnetic phase transitions \cite{Cao15} and quasiparticles \cite{Cao17}. 
Based on this simulation scheme, we show that the monopole can exert an attractive interaction onto the kink through the
monopolar magnetic field, which gives rise to the binding of the two quasiparticles.
The binding can also be released by a kink-antikink annihilation. 
In essence, our simulation scheme comprises coupling effects of the magnetic monopole to
other magnetic quasiparticles and reveals their binding and antibinding transition.
It provides a tool to control and manipulate the dynamics of magnetic monopoles.

	\begin{figure*}[t]
		\includegraphics[trim=65 10 65 10,width=0.80\textwidth]{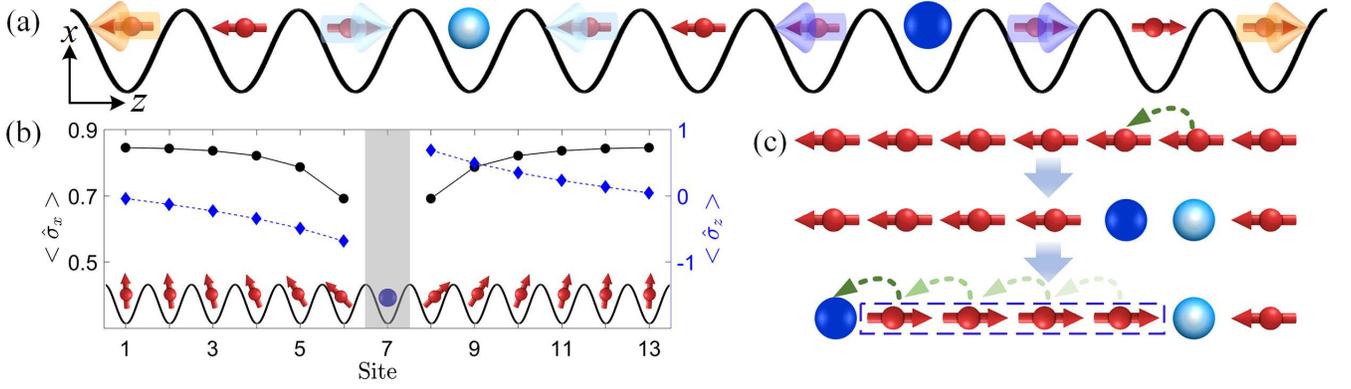}
		\caption{\label{fig:epsart} (a) The pseudospin chain based on the DSG system.
		The NM and SM are sketched with the dark and light blue balls, respectively.
		The transparent arrows of orange, dark and light blue color refer to the ABM and the monopolar
		magnetic fields around the NM and SM, respectively. (b) The polarization of neighboring spins
		around a localized NM, in terms of $\left\langle \hat \sigma_x \right\rangle$ (solid lines) and
		$\left\langle \hat \sigma_z \right\rangle$ (dashed lines). The spin polarization is also 
		explicitly shown at the bottom. (c) The dynamical process of the pair excitation, 
		the tunneling of the NM and SM, as well as the spin flipping along the tunneling are shown.}
		\end{figure*}

This paper is organized as follows. 
In Sec. \ref{section:II}, we demonstrate the pseudospin chain based on the dipolar superlattice gas.
In Sec. \ref{section:III}, we focus on the interaction effects between the north monopole and kink.
A brief discussion and overlook are given in Sec. \ref{section:IV}.

\section{SETUP AND PSEUDOSPIN MAPPING} \label{section:II}
We consider the dipolar superlattice gas (DSG) of spin polarized fermions confined in the one-dimensional double-well superlattice, in which the fermions interact with each other through the repulsive dipole-dipole interaction (DDI) \cite{Lewe09}. 
The DSG system can be described by the following Fermi-Hubbard Hamiltonian:
\begin{equation}
\begin{aligned}
{\hat H_{\rm{FH}}}&=  - J\sum\limits_{i = 1}^M {\left( {\hat f_{2i}^\dag {\hat f_{2i - 1}} + {\rm{H}}.{\rm{c}}.} \right)} 
 - {J_1}\sum\limits_{i = 1}^{M - 1} {\left( {\hat f_{2i}^\dag {{\hat f}_{2i + 1}} + {\rm{H}}.{\rm{c}}.} \right)}\\
 & + \sum\limits_{i < j \in \left[ {1,2M} \right]} {{V_d}\left( {j - i} \right){{\hat n}_i}{{\hat n}_j}},\label{Eq:1}
\end{aligned}
\end{equation}
where  $\hat f_{2i-1/2i}^\dag$ ($\hat f_{2i-1/2i}$) is the fermionic creation (annihilation) operator at the left/right site of the $i$-th supercell, 
and the operator  ${\hat n_i} = \hat f_i^\dag {\hat f_i}$ counts the number of fermions at site $i$. 
The first two terms in ${\hat H_{\rm{FH}}}$  describe the intra- and inter-cell hopping, respectively, 
with the hopping amplitudes $J \gg {J_1}$ . The DDI between two fermions located in the $i$- and $j$-th site
is taken as ${V_d}\left( {j - i} \right) = d/\left| {{x_j} - {x_i}} \right|^{3}$, where ${x_i}$(${x_j}$) is the local minimum of the corresponding site, 
and $d$ denotes the DDI strength. 
Without loss of generality, we take $J = 10{J_1}$  and $\left( {{x_{2j}} - {x_{2j - 1}}} \right)/\left( {{x_{2j + 1}} - {x_{2j}}} \right) = 1/2$, 
with the lattice constant $a=x_{2j+1}-x_{2j-1}$. 
We further truncate the DDI to the nearest neighbor interaction, 
which presents a good approximation for the parameter regime explored in this paper.

The pseudospin mapping transfers the DSG system to an effective spin chain, and we generalize it and bring in defects to the spin chain.
Under the tight-binding approximation, each cell of the DSG system accommodates four local occupation states of $\left\{ {{{\left| {1,0} \right\rangle }_i},{{\left| {0,1} \right\rangle }_i},{{\left| {1,1} \right\rangle }_i},{{\left| {0,0} \right\rangle }_i}} \right\}$, 
where ${\left| {{n_L},{n_R}} \right\rangle _i}$ denotes ${n_L}/{n_R}$ fermions occupying the left/right site of the \textit{i}-th cell. In the pseudospin mapping,
the single-occupation states ${\left| {1,0} \right\rangle _i}$ and ${\left| {0,1} \right\rangle _i}$ are mapped to the spin states ${\left|  \leftarrow  \right\rangle _i}$ and ${\left|  \rightarrow  \right\rangle _i}$ at the \textit{i}-th site of the chain. 
We further map the double-occupation  ${\left| {1,1} \right\rangle _i}$ and local vacuum state ${\left| {0,0} \right\rangle _i}$ to the particle and hole defects of the spin chain, denoted by ${\left| P \right\rangle _i}$  and ${\left| H \right\rangle _i}$, respectively. 
Under the pseudospin mapping, the DSG system is mapped to an effective spin chain,
with the Hamiltonian ${\hat H^{\rm{DSG}}_{\rm{spin}}} = {\hat H_0} + {\hat H_{{\rm{SD}}}}$:
\begin{equation}
\begin{aligned}
{\hat H_0} &=  - J\sum\limits_{\alpha=1} ^M {\hat \sigma _x^\alpha }  - \frac{d}{4}\sum\limits_{\alpha  = 1}^{M - 1} {\hat \sigma _z^\alpha \hat \sigma _z^{\alpha  + 1}}  + \frac{d}{2}\sum\limits_{\alpha=1} ^M {\tilde \sigma _z^\alpha }\\
 &+ \frac{d}{4}\left( {\hat \sigma _z^1 - \hat \sigma _z^M} \right) + \frac{d}{4}\left( {\tilde \sigma _z^1 + \tilde \sigma _z^M} \right), \label{Eq:1}
\end{aligned}
\end{equation}

\begin{equation}
\begin{aligned}
{{\hat H}_{{\rm{SD}}}} &= \frac{d}{4}\sum\limits_{\alpha=1} ^M {\tilde \sigma _z^\alpha \left( {\hat \sigma _z^{\alpha  - 1} - \hat \sigma _z^{\alpha  + 1}} \right)} \\
&- {J_1}\sum\limits_{\alpha=1} ^{M - 1} {\left[ {\left( {\hat s_{ \to ,H}^\alpha \hat s_{H, \leftarrow }^{\alpha  + 1} + H.c.} \right) + \left( {\hat s_{ \leftarrow ,P}^\alpha \hat s_{P, \to }^{\alpha  + 1} + H.c.} \right)} \right]} \\
&- {J_1}\sum\limits_{\alpha=1} ^{M - 1} {\left[ {\left( {\hat s_{H, \to }^\alpha \hat s_{P, \to }^{\alpha  + 1} + H.c.} \right) + \left( {\hat s_{P, \leftarrow }^\alpha \hat s_{H, \leftarrow }^{\alpha  + 1} + H.c.} \right)} \right].} \label{Eq:1}
\end{aligned}
\end{equation}
In ${\hat H_0}$ , $\hat \sigma _x^{\alpha}$  and  $\hat \sigma _z^{\alpha}$  are the Pauli operators exerted to the pseudospins, and $\tilde \sigma _z^{\alpha} \equiv {\left| P \right\rangle _{\alpha}}\left\langle P \right| - {\left| H \right\rangle _{\alpha}}\left\langle H \right|$ 
is the effective Pauli operator exerted to the defects.
To describe the counter-propagation of the defect and pseudospin, the exchanging operator
is introduced in ${\hat H_{\rm{SD}}}$, 
defined as $\hat s_{\Lambda ,\sigma }^{\alpha} \equiv {\left| \Lambda  \right\rangle _{\alpha}}\left\langle \sigma  \right|$ , with $\Lambda \in \left\{{P,H} \right\}$  and $\sigma \in \left\{ { \to , \leftarrow } \right\}$ . 
In the following, we term the effective spin chain as the DSG pseudospin chain.

The DSG pseudospin chain is manifested as a transverse Ising spin chain, of which the pseudospins interact by the Ising-type spin-spin interaction and are subjected to a transverse magnetic field, as indicated by the first two terms of ${\hat H_0}$. 
The fourth term  of ${\hat H_0}$ refers to the antiparallel boundary magnetic (ABM) field localized at the two edges of the chain. 
The ABM field has been recognized as an efficient way to excite kink since the original studies on the magnetic kinks \cite{Wres95,Sch77,Macris00,Sara04}, 
and its gives rises to an intrinsic kink in the DSG pseudospin chain. More interestingly, 
as indicated by the last term of ${\hat H_0}$, the ABM fields also exert an attractive/repelling potential to the hole/particle defect along the direction of the field, 
which mimics the response of the magnetic south/north monopole to the external magnetic field.

\begin{figure}[t]
\includegraphics[trim=15 5 2 0,width=0.46\textwidth]{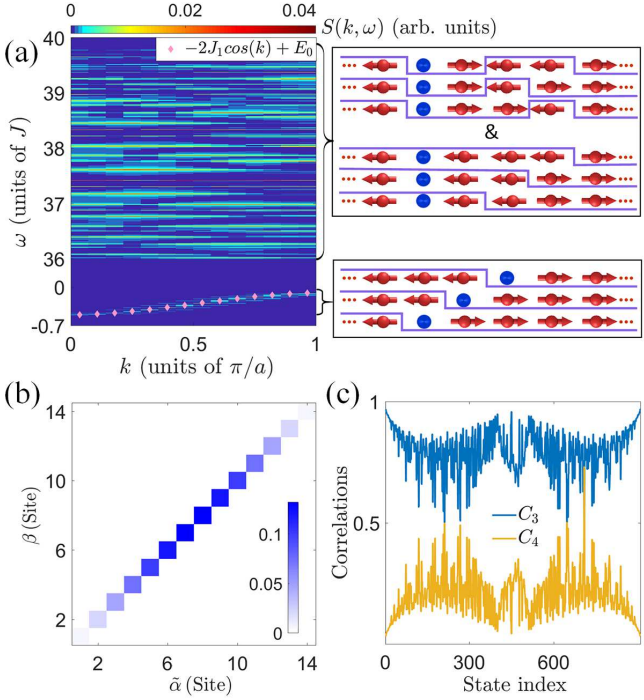}
\caption{\label{fig:epsart} (a) The dynamical structure factor $S(k,\omega)$ for the 14-sites pseudospin chain with $d=40J$. 
The frequency interval of $\omega \in (0.5,36)$ is removed, where the gap between the first two bands lies.
The representative basis states contributed to each band are shown to the right of the structure factor figure, 
in which different ferromagnetic domains are emphasized with solid-line steps and the (anti)kinks locate at the edges of the steps. 
The pink diamond in the main figure are the band dispersion obtained from the effective single-particle Hamiltonian describing the emergent particle composed of the monopole and kink.
(b) The NM-kink correlation for the ground state. 
(c) The three-body correlation $C_3$ (blue) and four-body correlation $C_4$ (yellow) of the second band.}
\end{figure}

Besides the response to the magnetic field, further fingerprints of the magnetic monopole,
i.e. the monopolar magnetic field and the Dirac string are also reproduced by the particle and
hole defects, which enables the particle and hole defects well resemble 
the north monopole (NM) and south monopole (SM), respectively. The monopolar magnetic field is normally evidenced by the spin texture around the monopole, and as indicated by the first term of ${{\hat H}_{{\rm{SD}}}}$, 
the particle (hole) defect polarizes the neighbor spins away from (towards) the defect, which resembles the monopolar magnetic field surrounding the NM (SM). The second term of ${{\hat H}_{{\rm{SD}}}}$ further demonstrates that the hopping of the defects is accompanied by the flipping of the counter-propagating spin, 
which has been recognized as a signature of the Dirac string for the monopoles \cite{Sondhi08,Hold09,Bram09,Perry09}.
The last term in ${{\hat H}_{{\rm{SD}}}}$ refers to the pair production of an NM and a SM, which manifests as the main excitation channel of the monopoles in spin ices.

In Fig. 1(a) the DSG pseudospin chain is sketched, in which the ABM and the monopolar magnetic field around the NM and SM is schematically shown. 
Figure 1(b) shows the expectation values of $\left\langle \hat \sigma_z \right\rangle$ and
$\left\langle \hat \sigma_x \right\rangle$ of the pseudospins around
a localized NM in the paramagnetic phase,
in which the transverse magnetic field aligns the pseudospins to the $x$-direction.
It can be seen that, the pseudospins far away
from the NM aligns along the transverse direction, while the neighboring spins to the NM are
polarized away from the NM, as indicated by the spin texture at the bottom of the figure.
 Figure 1(c) summarizes the pair production and spin-flipping effects of 
${{\hat H}_{{\rm{SD}}}}$  with a dynamical process that initially a pair of monopoles are excited and then hop away from each other, accompanied by the spin flipping. 
The detailed sketch of the pseudospin mapping and the comparison between the spin polarization around the NM, SM and a
normal magnetic defect in the pseudospin chain are given in Appendix A and B, respectively.

\section{INTERACTION EFFECTS BETWEEN NM AND KINK} \label{section:III}
In the strong interaction regime, the DSG pseudospin chain sustains the coexistence of the magnetic monopole and kinks, and provides an ideal platform to investigate the interplay of the two quasiparticles. 
Here, we focus on the doping of a single NM defect to the DSG chain, and the results can be straightforwardly generalized to the SM doping. 
We define the tail-to-tail and head-to-head kink as the kink and antikink, respectively. 
The Hilbert space is truncated to the subspace spanned by the basis states ${\left| n \right\rangle _{NM}} \otimes \left| {{{\tilde \alpha}_0},{{\tilde \alpha}_1} \cdots {{\tilde \alpha}_{2N}}} \right\rangle$,
in which ${\left| n \right\rangle _{NM}}$ denotes the position of the monopole and   
$\left| {{{\tilde \alpha}_0},{{\tilde \alpha}_1} \cdots {{\tilde \alpha}_{2N}}} \right\rangle  = \left| { \cdots  \leftarrow { \leftarrow _{{{\tilde \alpha}_0}-1}} \to  \cdots { \to _{{{\tilde \alpha}_1-1}}} \leftarrow  \cdots { \leftarrow _{{{\tilde \alpha}_{2N}-1}}} \to  \cdots } \right\rangle $ 
indicates the location of the (anti)kinks in the squeezed space where the monopole site is removed \cite{Timon17,Jayadev20}, 
(more details of the definition of the squeezed space is given in Appendix C). 
Accordingly, the Hamiltonian can be spanned in the monopole-kink subspace as:
$\hat H_{\rm{doped - spin}} = \hat H_{\rm {K}} + \hat H_{\rm{NM - K}}$, in which,

\begin{equation}
\begin{aligned}
{\hat H_{\rm{K}}} = d\sum\limits_{\tilde \alpha  = 1}^M {\hat n_A^{\tilde \alpha }}  - J\sum\limits_{\tilde \alpha  = 1}^{M - 1} {\left( {\hat S_ + ^{\tilde \alpha } + \hat S_ - ^{\tilde \alpha }} \right)\left( {\hat S_ + ^{\tilde \alpha  + 1} + \hat S_ - ^{\tilde \alpha  + 1}} \right)} ,
\end{aligned}
\end{equation}
\begin{equation}
\begin{aligned}
{\hat H_{\rm{NM - K}}} =  - d\sum\limits_{\alpha  = \tilde \alpha}^M {\hat n_K^{\tilde \alpha}\hat n_N^\alpha }  - {J_1}\sum\limits_{\alpha  = \tilde \alpha}^{M - 1} {\left( {\hat b{{_N^\alpha }^\dag }\hat b_N^{\alpha  + 1}\hat S_ - ^{\tilde \alpha }\hat S_ + ^{\tilde \alpha  + 1} + H.c.} \right)} .
\end{aligned}
\end{equation}
In ${\hat H_{\rm{K}}}$, $\hat n_{A}^{\tilde \alpha}$ refers to the number of antikinks between sites $\tilde \alpha$ and $\tilde \alpha+1$ in the squeezed space, and $\hat S^{\tilde \alpha \dag}_{+/-}  = (\hat a_{A}^{\tilde \alpha \dag}  + \hat a_{K}^{\tilde \alpha) }/{(\hat a_{A}^{\tilde \alpha}  + {\hat a_{K}^{\tilde \alpha \dag})}}$ 
combines the creation of a kink and the annihilation of a antikink. In ${\hat H_{\rm{NM - K}}}$ , $\hat b_{N}^{\alpha \dag} $  ( $\hat b_{N}^{\alpha}$) denotes the creation (annihilation) of a NM on the $\alpha$-th site, with ${\hat n_{N}^{\alpha}} = \hat b_{N}^{\alpha \dag} \hat b_{N}^{\alpha}$.  
${\hat H_{\rm{NM - K}}}$ then describes the interaction between the monopole and kinks, 
which includes the attractive interaction between a NM and a kink, and the effect of monopole hopping to the (anti)kink.

The interaction effects between the NM and the (anti)kinks can be captured by the dynamical structure factor $S(k,\omega)$
\cite{Grusdt18,Grusdt20}, and the $S(k,\omega)$ of the DSG pseudospin chain is shown in Fig. 2(a), which is calculated using  the multi-layer multi-configuration time-dependent Hartree method for 
arbitrary bosonic (fermionic) mixtures \cite{Schmelcher13,Cao13,Schmelcher17}, 
(for more details see Appendix D). 
In Fig. 2(a), a single-mode branch appears in the first band, and gives a strong hint that that the doped NM and
the intrinsic kink are bound and behave as a single composite quasiparticle.
The emergence of the NM-kink bound state can be confirmed by the 
NM-kink correlation $\left\langle \psi  \right|{\hat n_{K}^{\tilde \alpha}}{\hat n_{N}^{\beta}}\left| \psi  \right\rangle $ with
$\left| \psi  \right\rangle $ running through all eigenstates in the first band. 
Fig. 2(b) shows the NM-kink correlation for an arbitrary eigenstate in the first band, and it clearly demonstrate that
the NM and the intrinsic kink always occupy the same site. The NM-kink correlations for the other eigenstates in
the first band all present the same bound behavior, though not shown here.

\begin{figure}[t]
\includegraphics[trim=30 20 5 0,width=0.48 \textwidth]{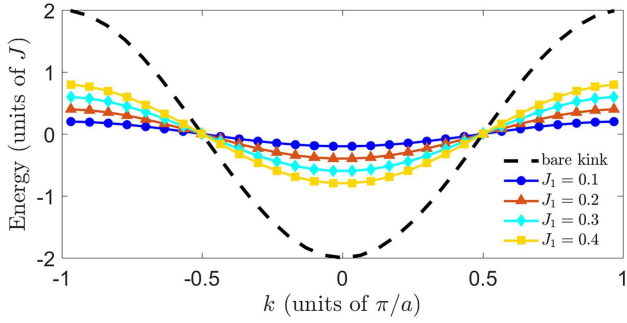}
\caption{\label{fig:epsart}
The dispersions of a bare kink (black dashed line) and the composite quasiparticle for $J_1=0.1$ (blue circle), $0.2$ (brown triangle), $0.3$ (cyan diamond) and $0.4$ (yellow square).}
\end{figure}

The second band in Fig. 2(a) presents a broad spectrum, leading to a continuum band in the infinite-long chain limit.
It is known that in the absence of the NM, the antikink-kink pair excitation dominates the excitation from the ground
to higher bands, which leads to continuum excited bands in the infinite-chain limit. 
An analysis via the multi-particle correlations, however, reveals that the presence of a NM not only preserves the
excitation channel of the antikink-kink pair excitation, but also brings in a new channel contributing
to the second band, which is the deconfinement of the NM and the intrinsic kink.
The multi-particle correlations have become powerful and experimentally accessible tools to identify
quasiparticle excitation in the ultracold-atom simulated pseudospin chains \cite{Demler21},
and here we determine the four-body correlation
$C_4=\sum\nolimits_{\alpha,\beta,\gamma} {\left\langle {\hat n_{K} ^{\alpha}
\hat n_{N} ^{\alpha}\hat n_{K}^{\beta} \hat n_{A}^{\gamma}} \right\rangle }$ and the three-body correlation
$ C_3= \sum\nolimits_{\alpha,\beta} {\left\langle {(1-\hat n^{\alpha}_K) 
\hat n^{\alpha}_N \hat n^\beta_K }\right\rangle}$ to identify the excitation channels from the first to the second band,
where $\alpha$, $\beta$ and $\gamma$ run over all sites in the (squeezed) chain with $\alpha \neq \beta$.
$C_4$ and $C_3$ signify the antikink-kink pair excitation in the presence of the bound NM-kink, and the deconfinement
of the bound NM-kink, respectively. 
The non-vanishing correlations $C_4$ and $C_3$ for the eigenstates in the second band, shown in Fig. 2(c),
demonstrate that both excitation channels contribute to the second band, and also suggest that the excitation
of an antikink-kink pair can be transformed to the deconfinement state of the NM-kink bond, which has potential
applications for manipulation of the monopoles and kinks.
In Fig. 2, the boxes to the right of Fig. 2(a) sketch the dominant contributions to the first two bands, 
and from bottom up they are the NM-kink bound state, the free pair of an NM and a kink and the coexistence of the bound NM-kink with the antikink-kink pair.

It is well known that composite quasiparticles composed of two types of particles, such as polarons \cite{Nagy18,Schm19}, can renormalize the dispersion and mobility of the bare particles and provides a unique control tool.
The NM-kink bound state also shares this renormalization effect to the bare kink. As shown in Fig. 3,  the dispersion of the bound state is significantly changed from that of the bare kink and can be tuned by the mobility of the NM.
Furthermore, to verify the manipulation of the bound NM-kink by the antikink-kink pair excitation,
we determine the dynamical process with the initial state that
a bound NM-kink and antikink-kink pair are located at the left and right edges of the pseudospin chain.
The temporal evolutions of $C_3(t)$ and $C_4(t)$ as shown in Fig. 4(a) indicate that, 
in the beginning of the dynamics the system is dominated by the coexistence of the bound NM-kink and the antikink-kink pairs,
whereas for later times as marked by the grey vertical lines in the figure, the deconfinement of the bound NM-kink  takes place,
accompanied with the disappearance of the coexistence of the NM-kink and antikink-kink pairs.
This confirms that the antikink-kink pair excitation can induce the deconfinement through the annihilation of the antikink
with the kink bound to the NM.

 \begin{figure}[t]
	\includegraphics[trim=30 10 30 0,width=0.42\textwidth]{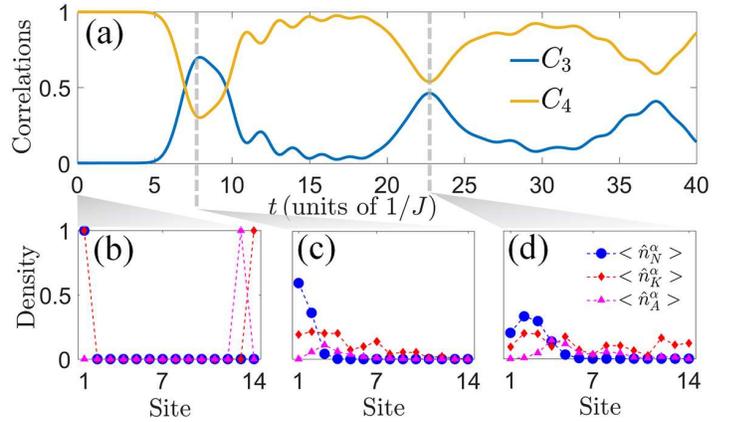}
	\caption{\label{fig:epsart} (a) Temporal evolution of $C_3$ (blue) and $C_4$ (yellow) during the dynamical process. 
	(b-d) The spatial densities of the NM $\left\langle {\hat n^{\alpha}_N} \right\rangle$ (blue circle), 
	the kink $\left\langle {\hat n^{\tilde \alpha}_K} \right\rangle$ (red diamond) and the antikink $\left\langle {\hat n^{\tilde \alpha}_A} \right\rangle$ (purple triangle) at the beginning (b) and later times (c-d) marked by grey vertical lines in (a).}
	\end{figure}

In Fig. 4(b)-(d), we also show the spatial densities of the NM, kink and antikink at the beginning and later times
marked in Fig. 4(a). The densities clearly show that in the beginning the NM-kink and antikink-kink pairs are separately
located on opposite edges, and at later times the NM and kink become deconfined with the antikink almost vanished.
The spatial densities further verify the deconfinement of the bound NM-kink by the antikink-kink annihilation process.
The renormalization effect and the manipulation of the NM-kink bound state with the antikink
suggests rich interaction effects between the NM and the (anti)kinks, 
and provides a potential control manner of the kink by monopoles.

\section{DISCUSSION AND OVERLOOK} \label{section:IV}
Ultracold atoms have become an important platform for quantum simulation
and allow to implement various atomic pseudospin models. The latter enabled the simulation of
different magnetic quasiparticles, such as the magnons \cite{Gross13}, spin/magnetic polarons \cite{Greiner19,Koep19},
spinons \cite{Jayadev20}, kinks \cite{Cao15,Gorshkov19,Monroe21} as well as the monopole \cite{Piet0902,Piet0906,Ha14,Ha15,Spielman18}. 
The existing simulations mainly focus on the 
excitation condition and the dynamical properties of quasiparticles of an individual type. 
The DSG pseudospin system allows however for the coexistence and coupling i.e.
interaction effects of the magnetic monopole
and kink, which enriches the previously investigated scenario of the quantum simulation of individual
magnetic quasiparticles with ultracold atoms.

The key ingredients of the DSG pseudospin scheme involve the double-well superlattice and the dipolar interaction,
which are realizable within the current experimental techniques.
The double-well superlattice is typically realized by the superposition of two pairs of counterpropagating laser beams \cite{Sebby06,Bloch07,Brown07,yuan16},
with $\lambda_1=2\lambda_2$, where $\lambda_{1(2)}$ refer to the wavelengths of the laser beams. 
The dipolar quantum gases can be composed of ultracold polar atoms \cite{Lee17,Ollikainen17}, 
Rydberg atoms \cite{Wuster15,Nguyen18} and polar molecules \cite{Zoller06,jun21}.
Particularly, our numerical simulations truncated the dipolar interaction to the nearest-neighbor interaction,
which can be implemented by e.g. the Rydberg dressing \cite{Gross16_r,Gross17_r}.
(An estimation of the experimental parameters is given in Appendix E.)
Moreover, this simulation scheme can be directly generalized to two-dimensional superlattice potentials,
which not only generalizes the spin chain to the two-dimensional square \cite{Pan20} and triangular 
lattices \cite{Sebby06, Porto07}, but also enables the simulation of the Dzyaloshinskii–Moriya-like
spin-spin interactions by exploring the anisotropy of the dipolar interaction.

Based on our simulation scheme, we have revealed binding and antibinding effects between the monopole 
and the kink. These effects are not restricted to the case of ultracold atomic pseudospins, but can be generalized to
the condensed matter spin systems. 
It is interesting to notice that a very recent experimental work investigating CoTb
films \cite{CoTb} reported the excitation of magnetic monopole pairs, in which the
excited monopole pairs are bound to a ferromagnetic domain wall, i.e. the two-dimensional counterpart 
of the magnetic kink. The binding effect in both the pseudospin and condensed matter spin systems
can be attributed to the
common nature of the singular magnetic field of the monopole exerted onto neighboring spins,
which induces the attractive interaction between the monopole and the kink/domain wall. 
It can also be expected that the simulation based on ultracold quantum gases
would stimulate related investigations in (artificial) spin lattices \cite{Lopez21,Farhan19}.

\section*{ACKNOWLEDGMENTS}
The authors would like to acknowledge T.Shi and Y. Chang for inspiring discussions, as well as
Y. Cai for the helpful discussion on experimental realization of Rydberg dressing. 
This work was supported by the National Natural Science Foundation of China 
(Grants Nos. 11625417, No. 11604107, No. 91636219 and No. 11727809) and 
the Cluster of Excellence' Advanced Imaging of Matter' of the 
Deutsche Forschungsgemeinschaft (DFG)- EXC 2056 - project ID 390715994.

\begin{appendix}

\section{THE PSEUDOSPIN MAPPING}

In this section, a more visualizable demonstration of the pseudospin mapping is provided in Fig. 5. 
As introduced in the main text, there are four single-cell occupation states  $\left\{ {{{\left| {1,0} \right\rangle }_i},{{\left| {0,1} \right\rangle }_i},{{\left| {1,1} \right\rangle }_i}, {{\left| {0,0} \right\rangle }_i}} \right\}$, 
each of which is mapped to a spin state and/or defect state. 
The mapping of the four occupation states to the corresponding spin and/or defect states is now given in Fig. 5(a). 
Following the pseudospin mapping, the whole superlattice loaded with atoms can be mapped to a spin chain with doped NM and SM, 
and the mapping between the superlattice of a particular atom filling configuration and the corresponding doped spin chain is shown in the Fig. 5(b).

 \begin{figure}[htbp]
	\includegraphics[trim=10 0 0 0,width=0.48\textwidth]{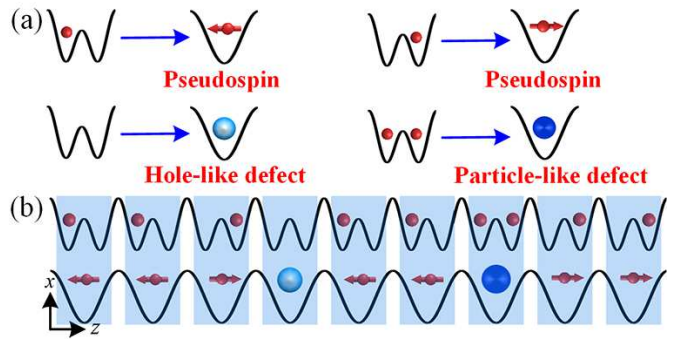}
	\caption{\label{fig:epsart} Illustration of the pseudospin mapping. (a) The occupation states of arbitrary cell are mapped to the pseudospin/defect states. (b) The original DSG system (upper panel) and the effective doped pseudospin chain (lower panel).}
	\end{figure}

\section{THE POLARIZATION EFFECT OF THE MONOPOLE}
Here, we provide more calculation results on the spin polarization effect of the NM and SM, which are compared to the spin polarization induced by a normal magnetic defect. To accomplish this, we consider a transverse Ising spin chain doped with a defect localized at the middle of the chain. Without the doping, the spins in the chain are all aligned to the x-direction by the transverse magnetic field, and the spin chain resides in the paramagnetic phase. The doped defect can interact with its neighboring spins, and polarize these spins to a 'new' direction. Defects of different types can result in very different spin textures of the neighboring spins. Here we separately consider three types of defects, namely the NM, SM and a normal magnetic defect, and compare the spin textures from these defects. We model the normal magnetic defect as a particle of 1/2 spin, and the spin state of the defect is fixed to ${\left|  \uparrow  \right\rangle _z}$.

\begin{figure*}[htbp]
\includegraphics[trim=40 10 40 0,width=0.9\textwidth]{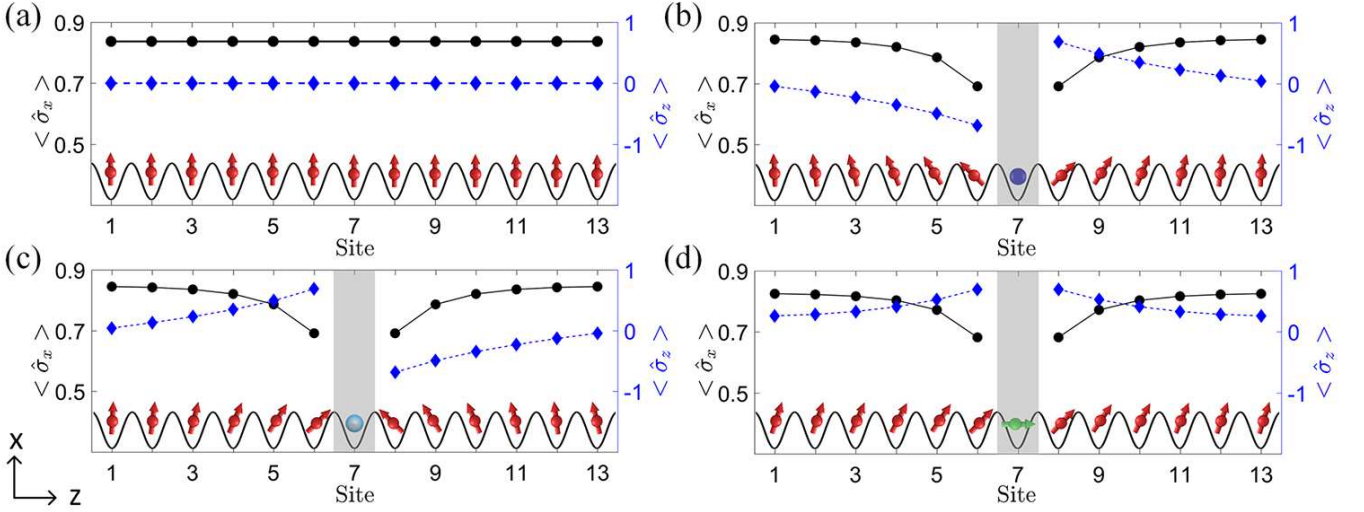}
\caption{The polarization effect for different defects for $d=3J$. (a) The magnetization along the x (black) and z (blue)-axis of the undoped spin chain, while (b)/(c) and (d) show the NM/SM and normal magnetic defect case. The black solid circles and blue squares represent $\left\langle {{{\hat \sigma }_x}} \right\rangle $ and $\left\langle {{{\hat \sigma }_z}} \right\rangle $, respectively.}
\end{figure*}

In our study, the Hamiltonian is taken as $\hat H = {\hat H_0} + \hat H_{{\rm{defect}}}^\alpha $, in which ${\hat H_0}$ refers to that of the transverse Ising spin chain, as introduced in the main text. $\hat H_{{\rm{defect}}}^\alpha $ corresponds to the spin-defect interaction, with $\alpha$=NM, SM and normal, denoting the NM, SM and normal magnetic defects. The different spin-defect interactions read:

\begin{subequations}
	\begin{align}
	&\hat H_{{\rm{defect}}}^{{\rm{NM}}} = \frac{d}{4}\sum\limits_\alpha ^M {\hat n_N^\alpha \left( {\hat \sigma _z^{\alpha  - 1} - \hat \sigma _z^{\alpha  + 1}} \right)},\\
	&\hat H_{{\rm{defect}}}^{{\rm{SM}}} = \frac{d}{4}\sum\limits_\alpha ^M {\hat n_S^\alpha \left( { - \hat \sigma _z^{\alpha  - 1} + \hat \sigma _z^{\alpha  + 1}} \right)} ,\\
	&\hat H_{{\rm{defect}}}^{{\rm{normal}}} = \frac{d}{4}\sum\limits_\alpha ^M {\hat n_{{\rm{nor}}}^\alpha \left( { - \hat \sigma _z^{\alpha  - 1} - \hat \sigma _z^{\alpha  + 1}} \right)}.
	\end{align}
	\label{eq}
\end{subequations}
$\hat H_{{\rm{defect}}}^{{\rm{normal}}}$ is taken from the situation of, for instance, doping magnetic atoms to electron gases, where the doped atom plays the role of the normal magnetic defect.
The results are shown in Fig. 6, and Fig. 6(a) shows the spin polarization of the undoped DSG spin chain in the weak interaction regime ($d=3J$), which correspond to the paramagnetic phase. In Fig. S2 (b) and (c), we present the results for the NM and SM defects. Here we observe that the neighboring spins are pointing away from (towards) the NM (SM) defect. Figure S2(d) provides a comparison to the normal magnetic defect, which polarizes the neighbor spins on its left- and right-hand side to the same direction.

\section{THE SQUEEZED SPACE}
In this section, we use Fig. 7 to give a more explicit demonstration of the basis defined in the squeezed space. Fig S3(a) shows one of the basis states of the pseudospin chain with nine spins labeled by $\alpha  = 1 \sim 9$. There are two kinks and an antikink locate between site $\left( {2,3} \right)$, $\left( {8,9} \right)$ and $\left( {6,7} \right)$ . The basis with a doped particle-defect at the 5-th site is shown in Fig. 7(b). Subsequently, we remove the particle-defect from the pseudospin chain and the left spins (Fig. 7(c)) form the squeezed space. The spins on the right side of particle-defect are squeezed forward. The left spins in squeezed space are relabeled by  $\tilde \alpha  = 1 \sim 8$, where $\tilde \alpha=\alpha$ and $\tilde \alpha=\alpha+1$ for $\alpha<5$ and $\alpha>5$, respectively. In the squeezed space, the kinks and antikink are located between sites $\left( {2,3} \right)$, $\left( {7,8} \right)$ and $\left( {5,6} \right)$. The falling and rising edges of the purple solid line in Fig. 7(d) indicate the positions of kinks and antikink, respectively.

\begin{figure}[htbp]
\includegraphics[trim=0 15 0 0,width=0.49\textwidth]{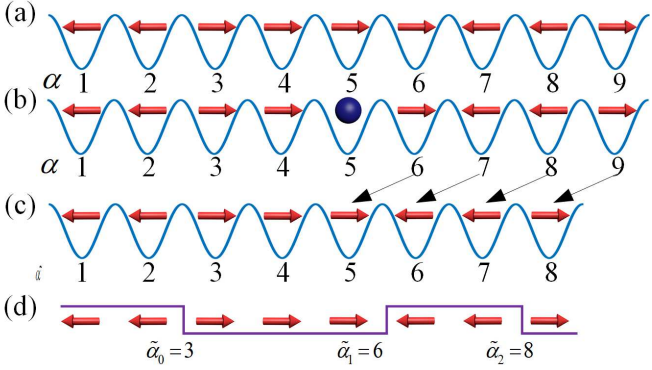}
\caption{Illustration of squeezed space. (a) The basis without defect. (b) The basis with single particle-defect at 5-th site. (c) The basis in squeezed space. (d) The kink-antikink basis.}
\end{figure}

\section{CALCULATIONS OF THE DYNAMICAL STRUCTURE FACTOR}
In this section, we present two approaches to calculate the dynamical structure factor. The first approach is based on [1], in which $S\left({k,\omega } \right)$ is determined using the Fourier transformation of the time-dependent correlation function. The second approach, following the proposal in \cite{Grusdt18}, couples the DSG system to a particle bath and applies a periodic variation of the coupling strength. Then $S\left( {k,\omega } \right)$ is extracted from the dynamical process under the periodic driving, which resembles the Angle-Resolved Photo Emission Spectroscopy (ARPES) signal. The two approaches give qualitatively the same results, and the main difference arises from the quantitative weights of each mode. Given the comparison between the two approaches, we show the results of the first approach in Fig. 2(a) in the main text.

Firstly, we provide the definition of the single-particle spectral function, which is the imaginary part of the single-particle retarded Green’s function. The single-particle retarded Green’s function ${G^R}\left( {k,t} \right)$ in Lehmann representation at zero temperature reads \cite{Damascelli04}:
\begin{equation}
\begin{aligned}
{G^R}\left( {k,\omega } \right) = \sum\limits_{n,\sigma } {\left\{ {\frac{{{{\left| {\left\langle {{\psi _n}\left| {\hat a_{k,\sigma }^\dag } \right|{\varphi _0}} \right\rangle } \right|}^2}}}{{\omega  + {\varepsilon _0} - {\varepsilon _n} + i\eta }} + \frac{{{{\left| {\left\langle {{\psi _n}\left| {{{\hat a}_{k,\sigma }}} \right|{\varphi _0}} \right\rangle } \right|}^2}}}{{\omega  - {\varepsilon _0} + {\varepsilon _n} + i\eta }}} \right\}},
\end{aligned}
\end{equation}
where $\hat a_{k,\sigma }^\dag  = \sqrt {2/\left( {M + 1} \right)} \sum\nolimits_{j = 1}^M {\sin \left[ {k \cdot j/\left( {M + 1} \right)} \right]\hat a_{j,\sigma }^\dag } $ creates a fermion with momentum $k$  and spin $\sigma$. $\left| {{\varphi _0}} \right\rangle $  is the ground state of the system without doping and  $\left| {{\psi _n}} \right\rangle $ is the  $n$-th eigenstate of the system with single particle doping, while ${\varepsilon _0}$  and  ${\varepsilon _n}$ are their energy. $\eta  \in {\mathbb{R}^ + }$ and we consider the limit $\eta  \to {0^ + }$. We focus on the single-particle excitations, and we therefore have ${\left| {\left\langle {{\psi _n}\left| {{{\hat a}_{k,\sigma }}} \right|{\varphi _0}} \right\rangle } \right|^2} = 0$. The single-particle spectral function $S\left( {k,\omega } \right)$ takes on the following appearance:
\begin{equation}
\begin{aligned}
{S^G}\left( {k,\omega } \right) &=  - \frac{1}{\pi }{\mathop{\rm Im}\nolimits} \left\{ {{G^R}\left( {k,\omega } \right)} \right\} \\
&= \sum\limits_{n,\sigma } {{{\left| {\left\langle {{\psi _n}\left| {\hat a_{k,\sigma }^\dag } \right|{\varphi _0}} \right\rangle } \right|}^2}\delta \left( {\omega  + {\varepsilon _0} - {\varepsilon _n}} \right)} ,
\end{aligned}
\end{equation}
and the density of states (DOS) is defined as ${\rho ^G}\left( \omega  \right) = \int {dk{\rm{ }}{S^G}\left( {k,\omega } \right)}$.

\begin{figure*}[htbp]
	\includegraphics[trim=10 20 10 10,width=0.9 \textwidth]{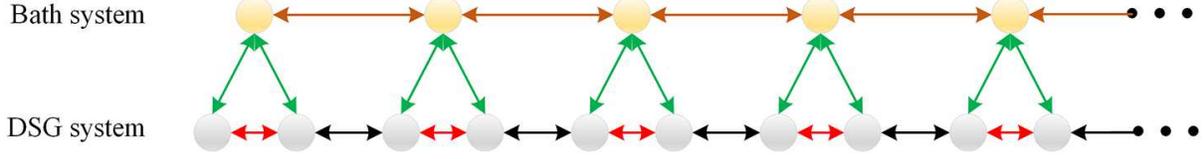}
	\caption{\label{fig:epsart} The proposed setup. The red (black) arrows indicate the intra-(inter-) cell hopping of spin polarized fermions in the DSG system. The brown arrows indicate the hopping of fermions in the bath, while the green arrow is the modulation between these two systems.}
	\end{figure*}

\subsection{Approach-1}
Following Eq. [A1] in Ref \cite{Grusdt20}, the approach firstly determines the spin-spin time-dependent correlation function, which is defined as:
\begin{equation}
\begin{aligned}
C\left( {i,t} \right) = \sum\limits_\sigma  {\left\langle {{\varphi _0}\left| {{e^{i\hat Ht}}{{\hat a}_{i,\sigma }}{e^{ - i\hat Ht}}\hat a_{1,\sigma }^\dag } \right|{\varphi _0}} \right\rangle }
\end{aligned}
\end{equation}
where $\hat a_{i, \to / \leftarrow }^\dag /{\hat a_{i, \to / \leftarrow }} = \hat f_{2i/2i - 1}^\dag /{\hat f_{2i/2i - 1}}$ is the fermionic creation (annihilation) operator at the right/left site of the $i$-th supercell. Then the spatial Fourier transform is performed and gives rise to
\begin{equation}
\begin{aligned}
A\left( {{{k}},t} \right) = \sqrt {\frac{2}{{M + 1}}} \sum\nolimits_i {\sin \left( {\frac{{i \cdot {{k}}}}{{M + 1}}} \right)C\left( {i,t} \right)}. 
\end{aligned}
\end{equation}
Finally, the dynamical structure factor  ${S^F}\left( {{{k}},\omega } \right)$  is obtained by the Fourier transformation in time:
\begin{equation}
\begin{aligned}
{S^F}\left( {{{k}},\omega } \right) = \frac{1}{{2\pi }}\int_{ - \infty }^\infty  {dtA\left( {{{k}},t} \right)},
\end{aligned}
\end{equation}
and the DOS is given by  ${\rho ^F}\left( \omega  \right) = \int {dk{\rm{ }}{S^F}\left( {k,\omega } \right)} $
\subsection{Approach-2}
The second approach involves the simulation of the ARPES process \cite{Grusdt18}. In the simulation of ARPES, the undoped DSG system is firstly connected to a bath system, which allows particle hopping between the DSG chain and the bath. A periodic shaking is then applied to the coupling strength, with a particular shaking frequency. During the dynamical process under the shaking, particles can tunnel from the bath to the DSG chain, corresponding to the doping of a particle to the DSG system. The structure factor is then associated with the momentum distribution of the hole in the bath at the particular driving frequency. We illustrate our simulation setup of this ARPES process as follows:
Firstly, we consider a (DSG+Bath) system as in Fig. 8, in which the bath is a lattice of atoms in the Mott state. The lattice of the bath system has the same period with the double-well superlattice but there is a single site per cell. We load the same spin polarized fermions into the DSG system and the bath system, while the DSG and bath system is half and unit filling, respectively.

The Hamiltonian of the (DSG+Bath) system reads:
\begin{subequations}
	\begin{align}
	&{\hat H_{{\rm{DSG}} + {\rm{bath}}}} = {\hat H_{{\rm{DSG}}}} + {\hat H_{{\rm{bath}}}} + {\hat H_{{\rm{intra}}}},\tag{D6} \\
	&{\hat H_{{\rm{DSG}}}} =  - J\sum\limits_{i = 1}^M {\left( {\hat f_{2i}^\dag {{\hat f}_{2i - 1}} + H.c.} \right)}  - {J_1}\sum\limits_{i = 1}^{M - 1} {\left( {\hat f_{2i}^\dag {{\hat f}_{2i + 1}} + H.c.} \right)}  \notag \\ 
	&+ \sum\limits_{i < j \in \left[ {1,2M} \right]}^{M - 1} {{V_d}\left( {j - i} \right){{\hat n}_i}{{\hat n}_j}}, \\
	&{\hat H_{{\rm{bath}}}} =  - {J_0}\sum\limits_{i = 1}^M {\left( {\hat c_i^\dag {{\hat c}_{i - 1}} + H.c.} \right)}  + \Delta \sum\limits_{i = 1}^{M - 1} {{{\tilde n}_i}}, \\
	&{\hat H_{{\rm{inter}}}} =  - {J_{{\rm{inter}}}}\sin \left( {{\omega _S}t} \right)\sum\limits_{i = 1}^M {\left[ {\hat c_i^\dag \left( {{{\hat f}_{2i - 1}} + {{\hat f}_{2i}}} \right) + H.c.} \right]},
	\end{align}
	\label{eq}
\end{subequations}

\noindent {where $\hat f_{2i/2i-1}^\dag /{\hat f_{2i/2i - 1}}$($\hat c_i^\dag /{\hat c_i}$) are the fermionic creation/annihilation operator at the right/left site of the $i$-th supercell (cell) in the DSG (bath) system, 
and the operator  ${\hat n_i} = \hat f_i^\dag {\hat f_i}$ (${\tilde n_i} = \hat c_i^\dag {\hat c_i}$) counts the number of fermions at site $i$ in the DSG (bath) system. The first two terms in ${\hat H_{{\rm{DSG}}}}$ describe the intra- and inter-cell hopping, respectively, 
with the hopping amplitudes $J$  and $J_1$. $\Delta$ is the offset of the bath relative to the DSG and the hopping amplitude of the fermions in the bath is $J_0$. 
The dipole-dipole interaction (DDI) between two fermions located at the $i$- and $j$-th site is taken as ${V_d}\left( {j - i} \right) = d/{\left| {{x_j} - {x_i}} \right|^3}$, 
where $x_i$ ($x_j$) is the local minimum of the corresponding site, and $d$ denotes the DDI strength. 
In the bath system, the fermions are well-separated from each other and we therefore ignore the DDI among them. 
The lattice modulation can be described by ${\hat H_{{\rm{inter}}}}$, ${J_{{\rm{inter}}}} \ll {J_1}$ is the perturbation term.}

${\hat H_{{\rm{inter}}}}$  induces hopping of atoms from the bath to the DSG system, and the energy change of the DSG system due to the doping is $\hbar \omega  = {E^{M + 1}} - {E^M}$ for single atom hopping. For a lattice modulation frequency $\omega_S$, this is determined by the energy conservation:
\begin{equation}
\begin{aligned}
\hbar \omega  = \hbar {\omega _S} - {E^B}\left( k \right) - \Delta ,
\end{aligned}
\end{equation}
where ${E^B}\left( k \right) =  - 2{J_0}\cos \left( k \right)$ is the energy of the hole in the bath system. The offset $\Delta$ is taken as $8d$ and $9d$ when we detect the dynamical structure factor of the first and second band, corresponding to the energy of a NM and the energy of one NM and antikink, respectively. The momentum and energy resolution spectrum function are obtained by detecting the momentum distribution of the hole for different $\omega_S$.

\begin{figure*}[t]
	\includegraphics[trim=30 10 20 10,width=0.9 \textwidth]{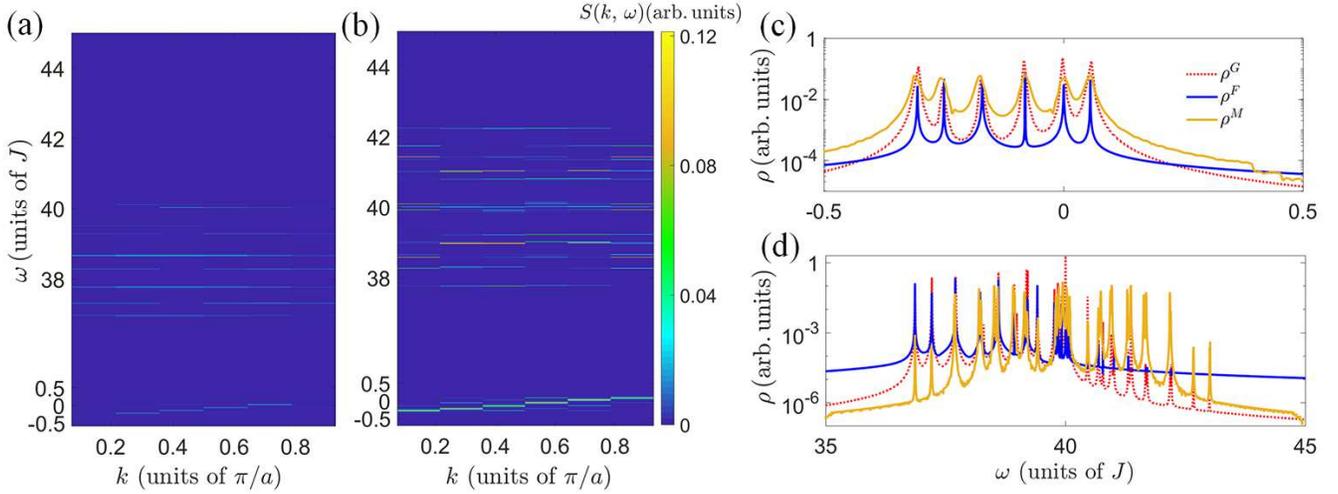}
	\caption{\label{fig:epsart} Dynamical structure factor of the 6-sites DSG pseudospin chain. (a) ${S^F}(k,\omega)$  obtained by the Fourier transform of the time-dependent correlation function. 
    (b) ${S^M}(k,\omega)$ originating from the lattice modulation. (c) and (d) are the DOS, with $\rho ^G$ (red dashed line), $\rho ^F$ (blue solid line), and $\rho ^M$ (orange solid line). }
	\end{figure*}

The structure factor ${S^M}\left( {k,{\omega _S}} \right)$ is determined from the reduced density matrix of the hole in the bath system as: 
\begin{equation}
\begin{aligned}
{S^M}\left( {k,{\omega _S}} \right) = \frac{\hbar }{{2\pi J_0^2}}\Gamma \left( {k,{\omega _S}} \right) ,
\end{aligned}
\end{equation}
where $\Gamma \left( {k,{\omega _S}} \right) = \frac{1}{{M + 1}}\sum\nolimits_{i,j} {\sin \left( {ik} \right)\sin \left( {jk} \right)} {R_{{\omega _S}}}\left( {i,j} \right)$ is the probability of creating a particle with momentum $k$ and energy $\hbar {\omega _S}$. ${R_{{\omega _S}}}\left( {i,j} \right) = {\rm{tr}_{{\rm{DSG}}}}\left[ {\left| {\psi \left( {{t_{{\omega _S}}}} \right)} \right\rangle \left\langle {\psi \left( {{t_{{\omega _S}}}} \right)} \right|} \right]$ is the reduced density matrix of the hole in the bath, while $\left| {\psi \left( {{t_{{\omega _S}}}} \right)} \right\rangle $ is the wavefunction of the complete system at $t_{\omega_S}$. At $t=t_{\omega_S}$, the probability of $\rm{tr}\left[ {{R_{{\omega _S}}}\left( {i,j} \right)} \right]$ hole takes a maximum for a given shaking with frequency $\omega_S$. The DOS is naturally defined as ${\rho ^M}\left( \omega  \right) = \int {dk{\rm{ }}{S^M}\left( {k,\omega } \right)} $.

In the strong interaction regime, which is of interest here, the gap between the adjacent bands is much larger than the strength of shaking. As a result, we only need to take the resonant states into account. For a certain $\omega_S$, after transforming to the rotating frame, performing a rotating wave approximation, one obtains the Hamiltonian:
\begin{subequations}
	\begin{align}
	&{\hat H^{{\rm{RWA}}}} = {\hat H_{{\rm{DSG}}}} + \hat H_{{\rm{bath}}}^{{\rm{RWA}}}{\rm{ + }}\hat H_{{\rm{inter}}}^{{\rm{RWA}}} ,\tag{D9} \\
	&{\hat H_{{\rm{DSG}}}} =  - J\sum\limits_{i = 1}^M {\left( {\hat f_{2i}^\dag {{\hat f}_{2i - 1}} + {\rm{H}}.{\rm{c}}.} \right)}  - {J_1}\sum\limits_{i = 1}^{M - 1} {\left( {\hat f_{2i}^\dag {{\hat f}_{2i + 1}} + {\rm{H}}.{\rm{c}}.} \right)}  \notag \\
	& + \sum\limits_{i < j \in \left[ {1,2M} \right]}^{M - 1} {{V_d}\left( {j - i} \right){{\hat n}_i}{{\hat n}_j}},\\
	&\hat H_{{\rm{bath}}}^{{\rm{RWA}}} =  - {J_0}\sum\limits_{i = 1}^M {\left( {\hat c_i^\dag {{\hat c}_{i - 1}} + {\rm{H}}.{\rm{c}}.} \right)}  + \left( {\Delta  - {\omega _S}} \right)\sum\limits_{i = 1}^{M - 1} {{{\tilde n}_i}},\\
	&\hat H_{{\rm{inter}}}^{{\rm{RWA}}} =  - \frac{{{J_{{\rm{inter}}}}}}{2}\sum\limits_{i = 1}^M {\left[ {\hat c_i^\dag \left( {{{\hat f}_{2i - 1}} + {{\hat f}_{2i}}} \right) + {\rm{H}}.{\rm{c}}.} \right]}.
	\end{align}
	\label{eq}
\end{subequations}

Throughout our numerical calculation, we set $J=1$ as the unit. The other parameters $J_0=J_1=0.1$, $d=40$, $J_{\rm{inter}}=0.01$ and the evolution time $T=1000$.

The structure factor computed with the first and second approach is given in Fig. 9(a) and (b) with  ${S^F}(k,\omega)$ and ${S^M}(k,\omega)$, respectively. We omit the internal of $\omega  \in \left( {0.5,35} \right)$ as ${S^{F/M}}(k,\omega) \sim 0$, which is the gap between the first two bands. ${S^X}(k,\omega)$ is normalized to unity $\int {d\omega {\rm{ }}{S^X}\left( {k,\omega } \right)}  = 1$ , with  $X=F,M,G$ \cite{Devereaux20}.
${S^F}(k,\omega)$ and  ${S^M}(k,\omega)$ are qualitatively the same, although it looks like that there are more details in Fig. 9(b). This is confirmed by the DOS shown in Fig. 5(c) and (d). Fig. 9(c) shows ${\rho ^G}(\omega)$ (purple), ${\rho ^F}(\omega)$ (blue) and ${\rho ^M}(\omega)$ (red) for $\omega  \in [ { - 0.5,0.5} ]$, and their peaks locate at the same position with similar amplitude. In Fig. 9(d), we show the DOS for $\omega  \in [ { - 35,45} ]$. The peaks are almost matched, although ${\rho ^F}(\omega)$ is invisible for the higher excited states of the second band. This is due to ${\left| {\left\langle {{\psi _n}\left| {\hat c_{k,\sigma }^\dag } \right|{\varphi _0}} \right\rangle } \right|^2}\sim 0 $ for these higher excited states. The most direct way to improve the intensity of ${S^F}( {k,\omega } )$ is taking the rest of the eigenstates of the undoped system into account.

The above analysis compares different approaches to obtain the dynamical structure factor, which give qualitatively the same spectrum, 
with the difference mainly arising in the quantitative amplitude of each mode. 
We then adapt the first approach since it is more setup-independent and not relying on the setting of e.g. the bath.

\section{THE EXPERIMENTAL REALIZATION}
Here, we discuss the experimental realizability of the DSG simulation scheme.
The key ingredients of our scheme involve the double-well superlattice and the dipolar interaction,
which are realizable within the current experimental techniques. 
The double-well superlattice is typically realized by the superposition of two pairs of counterpropagating laser beams \cite{Sebby06,Bloch07,Brown07,yuan16}.
The dipolar quantum gases can be composed of ultracold polar atoms \cite{Lee17,Ollikainen17}, 
Rydberg atoms \cite{Wuster15,Nguyen18} and polar molecules \cite{Zoller06,jun21}.
Particularly, our numerical simulations truncated the dipolar interaction to the nearest-neighbor interaction,
which can be implemented by e.g. the Rydberg dressing \cite{Gross16_r,Gross17_r}. The DSG simulation scheme also requires $U \gg J \gg J_1$, 
where $U$, $J$, and $J_1$ denote the strength of the NN interaction as well as the intra- and inter-cell hopping. 
In our numerical simulation, we take the parameters of $U=40J=400J_1$, where the DSG pseudospin chain resides to the single-kink phase. 

Taking $^{6}$Li atoms as the working medium, the wavelength of the laser beams to form the double-well superlattice are $\lambda_s=2.3$ $\mu$m and $\lambda_l=2\lambda_s$. 
Fixing the amplitudes of the lattice height of the short- and long-wavelength lattices as $V_s=18 E_{\rm{R}}$, 
$V_l=6.2 E_{\rm{R}}$ leads to the intra- and inter-cell hopping strength of $J=10J_1=68$ Hz, 
where  $E_{\rm{R}}=h^2/\left(2\lambda_s^2m_{\rm{Li}} \right)$ is the recoil energy, 
with $h$ and $m_{\rm{Li}}$ denoting the Planck constant and the atomic mass. 

\begin{figure}[h]
	\includegraphics[trim=10 10 0 0,width=0.46 \textwidth]{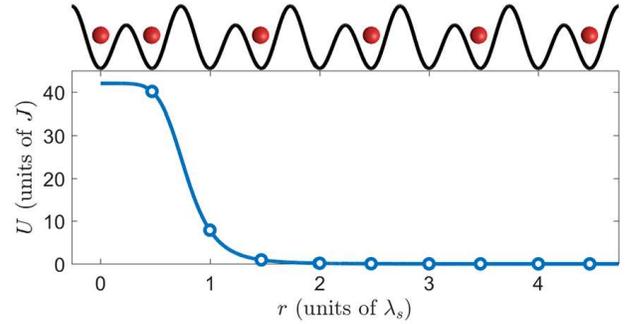}
	\caption{\label{fig:epsart} The interaction strength versus the relative distance between atoms, in units of $J$ and $\lambda_s$, respectively. 
	The top panel plots the double-well superlattice with the same length scale as relative distance in the main figure, 
	to demonstrate that the interaction mainly affects atoms in nearest neighbors.}
	\end{figure}

The NN interaction can be induced by the Rydberg dressing, 
and we take the Rydberg state as $\left. |34S_{1/2} \right \rangle$ for demonstration, 
of which the van-der-Waals type interaction coefficient $C_6=46.5$ MHz$\mu$m$^6$. 
To be consistent to the double-well superlattice settings, 
the NN interaction strength should take the value $U=2.72$ kHz, 
with the Rydberg radius approaching the period length of the lattice. 
It can be found that choosing the detuning and the Rabi frequency of the Rydberg excitation laser as $43.8$ MHz and $6.6$ MHz
 will leads to $U \sim 2.72$ kHz and Rydberg radius, $R_c=0.9$ $\mu$m, which meets the requirement of the setting of our numerical simulations. 
 We plot the interaction strength as a function of the relative distance between atoms in Fig. S10, 
 to visualize the NN interaction induced by the Rydberg dressing, and this interaction fulfills the DSG simulation scheme.

It is also worth mentioning that the simulation scheme is flexible with respect to the choice of parameters, and can be implemented over a wide parameter regime, 
enabling a feasible experimental realization.

\end{appendix}

\end{document}